\documentclass{article}
\usepackage[a4paper, total={6in, 8in}]{geometry}
\usepackage[sort&compress,numbers,square]{natbib}
\usepackage{orcidlink}
\usepackage{amsmath,amssymb,amsfonts}
\usepackage{algorithmic}
\usepackage{graphicx}
\usepackage{textcomp}
\usepackage{xcolor}
\usepackage{array}
\newcolumntype{N}{>{\centering\arraybackslash}m{1.0in}}
\newcolumntype{G}{>{\centering\arraybackslash}m{2in}}
\newcolumntype{P}[1]{>{\centering\arraybackslash}p{#1}}
\def\BibTeX{{\rm B\kern-.05em{\sc i\kern-.025em b}\kern-.08em
    T\kern-.1667em\lower.7ex\hbox{E}\kern-.125emX}}
\usepackage{listings}
\usepackage{caption}
\usepackage{subcaption}
\usepackage{enumitem}
\usepackage{tikz,lipsum}
\usepackage[most]{tcolorbox}
\usepackage{booktabs}
\usepackage{multirow}
\usepackage{makecell}
\usepackage{float}
\usepackage[normalem]{ulem}
\DeclareRobustCommand\full  {\tikz[baseline=-0.6ex]\draw[->, thick] (0,0)--(0.5,0);}
\DeclareRobustCommand\fulld  {\tikz[baseline=-0.6ex]\draw[<->, thick] (0,0)--(0.5,0);}

\DeclareRobustCommand\dottedd{\tikz[baseline=-0.6ex]\draw[<->, thick,dotted] (0,0)--(0.54,0);}
\DeclareRobustCommand\dashed{\tikz[baseline=-0.6ex]\draw[->, thick,dashed] (0,0)--(0.54,0);}

\hypersetup{pdfborder={0 0 0}}

\title{Model-based Digital Twins of Medicine Dispensers for Healthcare IoT Applications}

\author{Hassan Sartaj\orcidlink{0000-0001-5212-9787}$^1$ \and Shaukat Ali\orcidlink{0000-0002-9979-3519}$^{1,2}$ \and Tao Yue\orcidlink{0000-0003-3262-5577}$^1$ \and Kjetil Moberg\orcidlink{0009-0002-5042-7371}$^3$}

\date{
	$^1$Simula Research Laboratory, Oslo, Norway \\ \texttt{\{hassan, shaukat\}@simula.no}, \texttt{taoyue@gmail.com}\\%
    $^2$Oslo Metropolitan University, Oslo, Norway \\%
	$^3$Welfare Technologies Section, Oslo Kommune Helseetaten, Oslo, Norway \\ \texttt{kjetil.moberg@hel.oslo.kommune.no}\\[2ex]%
}

\begin{document}

\maketitle

\pagestyle{plain}

\begin{abstract}
Healthcare applications with the Internet of Things (IoT) are often safety-critical, thus, require extensive testing. Such applications are often connected to smart medical devices from various vendors. 
System-level testing of such applications requires test infrastructures physically integrating medical devices, which is time and monetary-wise expensive. Moreover, applications continuously evolve, e.g., introducing new devices and users and updating software. Nevertheless, a test infrastructure enabling testing with a few devices is insufficient for testing healthcare IoT systems, hence compromising their dependability.  
In this paper, we propose a model-based approach for the creation and operation of digital twins (DTs) of medicine dispensers as a replacement for physical devices to support the automated testing of IoT applications at scale.  
We evaluate our approach with an industrial IoT system with medicine dispensers in the context of Oslo City and its industrial partners, providing healthcare services to its residents. We study the fidelity of DTs in terms of their functional similarities with their physical counterparts: medicine dispensers.
Results show that the DTs behave more than 92\% similar to the physical medicine dispensers, providing a faithful replacement for the dispenser.  

\end{abstract}

\section{Introduction}  

An Internet of Things (IoT) healthcare application typically relies on a cloud-based infrastructure~\cite{balasubramanian2021scalable, das2022rescue}. 
Such infrastructure is the central access point for its stakeholders: healthcare professionals, caretakers, patients, etc.~\cite{gupta2021hierarchical}. 
In addition, smart medical devices are connected to the cloud infrastructure to provide services to patients.
Given the criticality of such applications, testing at various levels is needed to ensure their dependability. To enable system-level testing, a test infrastructure integrating many medical devices from different vendors with varied software applications is needed, which is, however, time-consuming, financially expensive, and practically infeasible, especially considering the continuous evolution of devices, software, stakeholders, etc.

The above-described challenge is faced by Oslo City's health department, which is responsible for providing a wide range of healthcare services to its residents, together with its industrial partners building medical devices and healthcare platforms to support these services. In this paper, we focus on medicine dispensers, devices provided to patients by Oslo City for timely deliveries of medicines, which are connected to various healthcare IoT applications provided by different industrial partners of Oslo City's health department.

To support the testing of a healthcare IoT application integrated with many medicine dispensers, we employ the digital twin (DT) technology~\cite{xu2023digital}. 
A DT is a virtual model of a real-world physical object~\cite{nath2021building}. 
DTs have been a key part of the industrial manufacturing process. 
DT technology has been successfully used in many domains, e.g., medical and healthcare domains~\cite{bruynseels2018digital,chengjie2023evoclinical}. 
In this paper, we use DTs to support large-scale testing of healthcare IoT applications by substituting physical medicine dispensers with their corresponding DTs.

To build DTs, we propose a model-based approach for the creation and operation of a DT representing a medicine dispenser. We first develop a domain model for medicine dispensers, with which an instance model representing the structure of a DT can be generated. Its behavior is modeled with an executable state machine, with which we can simulate the operation of a physical medicine dispenser with the DT. Moreover, to enable communications between DTs and a healthcare IoT application (the system under test (SUT)) and between DTs and their corresponding physical medicine dispensers, we developed the DT Communication Server. DTs and the server constructed with our approach can then be used as a system-level test infrastructure of a healthcare IoT application.

We evaluate our approach with an industrial case study provided by Oslo City for medicine dispensers, Karie~\cite{karie}. We analyze the fidelity of DT in terms of its functional similarities with a physical medicine dispenser. Results show that the DT functions more than 92\%, equivalent to a physical medicine dispenser. Results also show that our approach is scalable in integrating 100 DTs (simulating 100 physical medicine dispensers operating concurrently) into the test infrastructure. To summarize, our contributions are presented below.

\begin{itemize}

    \item A model-based approach to create and operate DTs of medicine dispensers. Our approach includes a domain model of medicine dispensers for structural modeling, executable state machines for behavioral modeling, integration with a healthcare IoT application, and communication and synchronization with physical medicine dispensers. 
    \item An empirical evaluation with a widely-used smart medicine dispenser and industrial healthcare IoT application. Our evaluation assesses the fidelity of DTs based on their functional similarity with physical medicine dispensers and the feasibility of simulating multiple devices with DTs. 
    \item An open-source implementation of our approach (available at GitHub~\cite{repo}) to facilitate further development and research.
  
\end{itemize}

% structure of the paper 
The rest of the paper is structured as follows. The real-world application context and challenges are discussed in Section~\ref{sec:iot-hc}. Our proposed model-based approach is presented in Section~\ref{approach}. The empirical evaluation of the proposed approach is demonstrated in Section~\ref{evaluation}. A discussion on results and lessons learned are presented in Section~\ref{lessonslearned}. The related works are described in Section~\ref{rws}. The paper is concluded in Section~\ref{conclusion}.

\section{Application Context}\label{sec:iot-hc}
In this section, we first discuss the real-world application based on which this study is conducted (Section \ref{subsec:real-world}). Then, we present the industrial context and challenges (Section \ref{subsec:industrialcontext}).

\subsection{Real-World Application} \label{subsec:real-world}
\begin{figure}[htbp]
\centerline{\includegraphics[width=7.4cm, height=7.6cm, keepaspectratio]{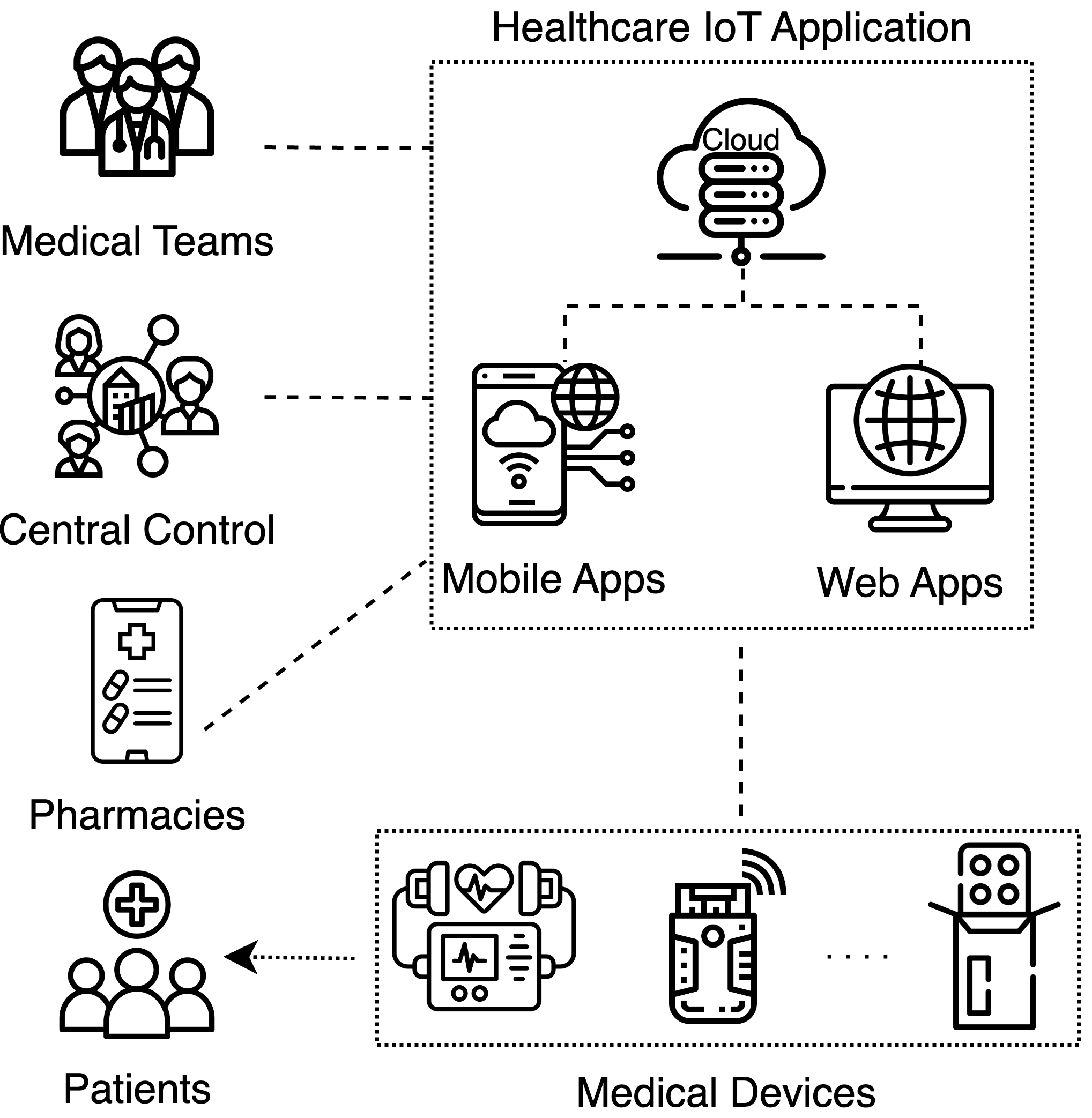}}
\caption{Context: an IoT-based healthcare system}
\label{fig:iot}
\end{figure}

The healthcare department of Oslo City provides healthcare facilities to its residents, including home care. 
Patients with different medical conditions require different solutions, e.g., dementia patients need medication at prescribed times, thereby requiring specialized machines for medicine reminders and delivery.
Thus, Oslo City uses modern IoT technologies to deal with varying patient requirements. For instance, a typical IoT-based healthcare system is shown in Figure~\ref{fig:iot}. The central part is a healthcare IoT application (composed of the IoT cloud, mobile applications, and web applications) connecting \textit{Central Control} (for access authorities), \textit{Patients}, \textit{Pharmacies}, and \textit{Medical Teams} (e.g., medical professionals from hospitals). The IoT cloud also connects smart medical devices such as automatic medicine dispensers, GPS trackers, and blood pressure/pulse measuring devices. 
Such devices are assigned to patients according to their health conditions and act as central elements in providing healthcare services, monitoring, and reporting health conditions via the IoT cloud to relevant stakeholders (e.g., nurses) when needed.

\subsection{Industrial Context and Challenges} \label{subsec:industrialcontext}
This work is conducted in the context of an innovation project with the healthcare department of Oslo City, focusing on automated testing of IoT-based healthcare services to significantly improve their quality, speed up the delivery of new services, and increase the solutions' scalability of handling an increasing number of patients and therefore devices. Oslo City works with several industrial partners to provide these services. One particular service is timely medicine delivery to patients via specialized dispensers, which is provided through an IoT-based healthcare system consisting of smart medical devices and cloud-based applications. They are interconnected to achieve the overall goal of providing high-quality services to relevant stakeholders, including the residents of Oslo City.

Our industrial healthcare IoT application has web and mobile applications for various stakeholders, e.g., Oslo City administration, patients, and nurses. 
This application supports integration with multiple smart medical devices of various types, such as medicine dispensers and health measuring devices, and enables bi-directional communication via API calls.  
A medicine dispenser provides the right medicine with the right amount at the right time to a patient and regularly reports the patient's medicine intake to the healthcare IoT application.
Medical professionals and nursing homes often use collected information for checking, adjusting, and analyzing patients' medication plans. 
These dispensers also notify concerned medical professionals in the case of emergency conditions through relevant applications, phones, etc., via APIs of the healthcare IoT application. 

Challenges faced by Oslo City and its industrial partners are 1) continuous and rapid addition of new services, e.g., integrating new devices cost-effectively and reliably into the overall system without affecting existing services; 2) evolving services, devices, and technologies, e.g., software updates; 3) serving more stakeholders which requires scalable solutions; and 4) high requirements on the availability and dependability of provided services. Many of these challenges directly affect the quality of services (QoS) Oslo City provides. Thus, in this project, we aim to ensure the QoS of the system with automated and rigorous testing of healthcare IoT applications with medical devices in the loop. 
To this end, we propose an approach to create and operate DTs of smart medicine dispensers to pave the way for automating system-level testing of IoT-based healthcare applications, saving required physical resources and time costs while ensuring the dependability of the SUT.

\section{Model-based DT Creation and Operation}\label{approach}
\begin{figure*}[!t]
\centerline{\includegraphics[width=\linewidth, height=\textheight, keepaspectratio]{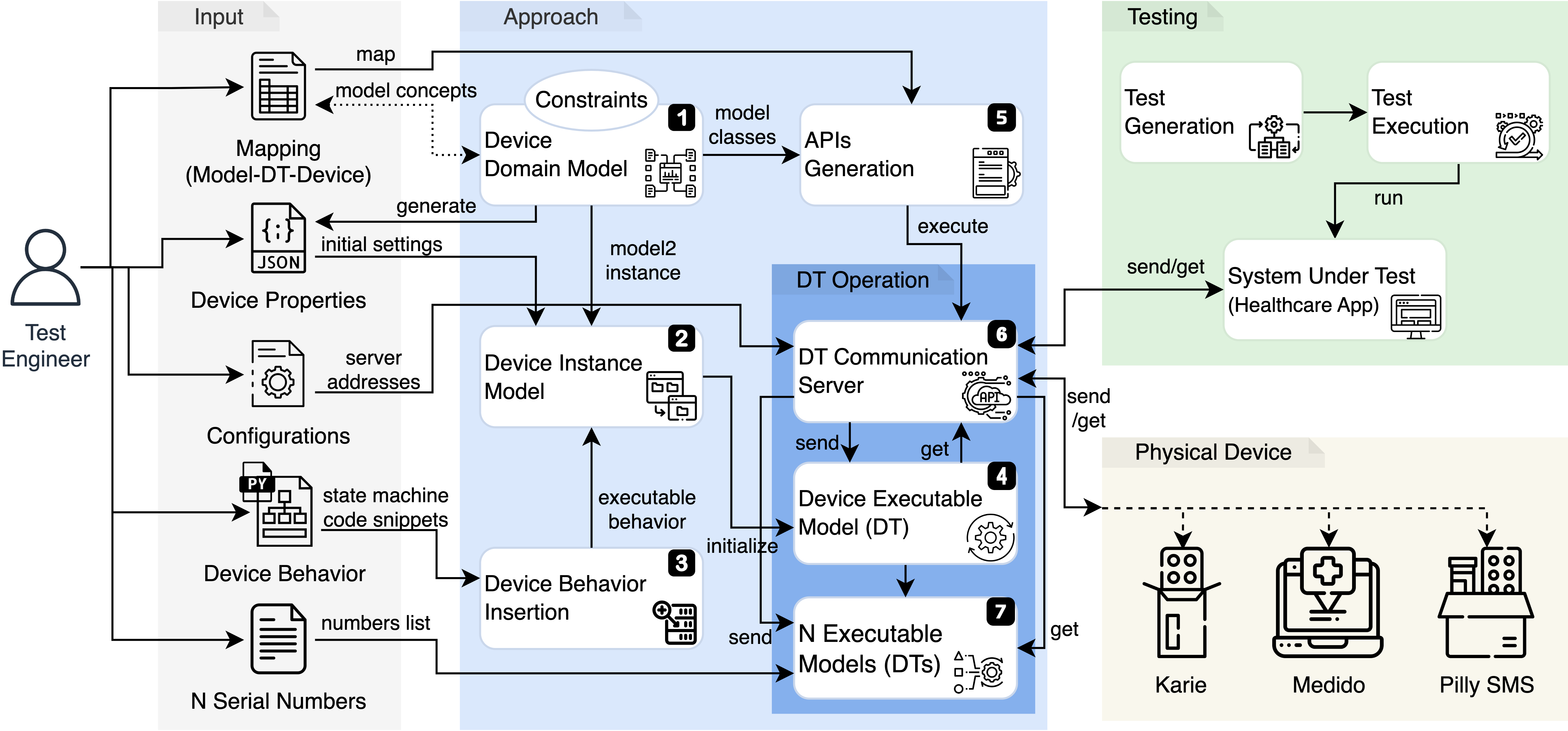}}
\caption{An outlook of our approach, including required inputs, its core components, integration with smart medicine dispenser devices, and testing components. The arrow (\full) shows information flow, (\fulld) shows two-way communication, (\dottedd) depicts a mapping, and (\dashed) represents communications with different medicine dispensers.}
\label{fig:app}
\end{figure*}

As shown in Figure~\ref{fig:app}, as the first step, our approach utilizes \textit{Device Domain Model} to generate an input template based on which device properties (e.g., brightness and volume) can be specified by \textit{Test Engineer}. Our approach takes these properties as input, populates the \textit{Device Domain model}, and automatically generates a \textit{Device Instance Model}, representing the structure of the DT we aim to develop. More details are presented in Section \ref{structural aspect of DT}.

The behavior of the DT is modeled as a state machine, which can either be given by test engineers or reused from an existing state machine built for the same or similar devices. Next, we link the behavior model (i.e., the state machine) with the DT's structural model (i.e., the instance model) and make them consistent. The behavior model is further appended with Python code to ensure it is executable such as the behavior of the devices (e.g., dispensing medicines) can be executed and hence simulated. Consequently, we obtain \textit{Device Executable Model}, as shown in Figure~\ref{fig:app}. More details are presented in Section \ref{behavioral aspect of DT}.

By taking the mapping between DT APIs and physical device APIs from test engineers, our approach generates DT APIs (\textit{APIs Generation}) based on the domain model. Our approach also configures \textit{DT Communication Server} based on configurations regarding server addresses from test engineers. It plays the role of connecting DTs with the SUT and with various types of physical devices via REST APIs. When everything is ready, the DT communicates with the SUT and optimally with the devices if needed. Additionally, the DT automatically validates its behavior when receiving new messages from the SUT and the devices. 
More details about how the DT operates are presented in Section \ref{operating DT}.

\begin{figure}[htbp]
\centerline{\includegraphics[width=\linewidth, height=\textheight, keepaspectratio]{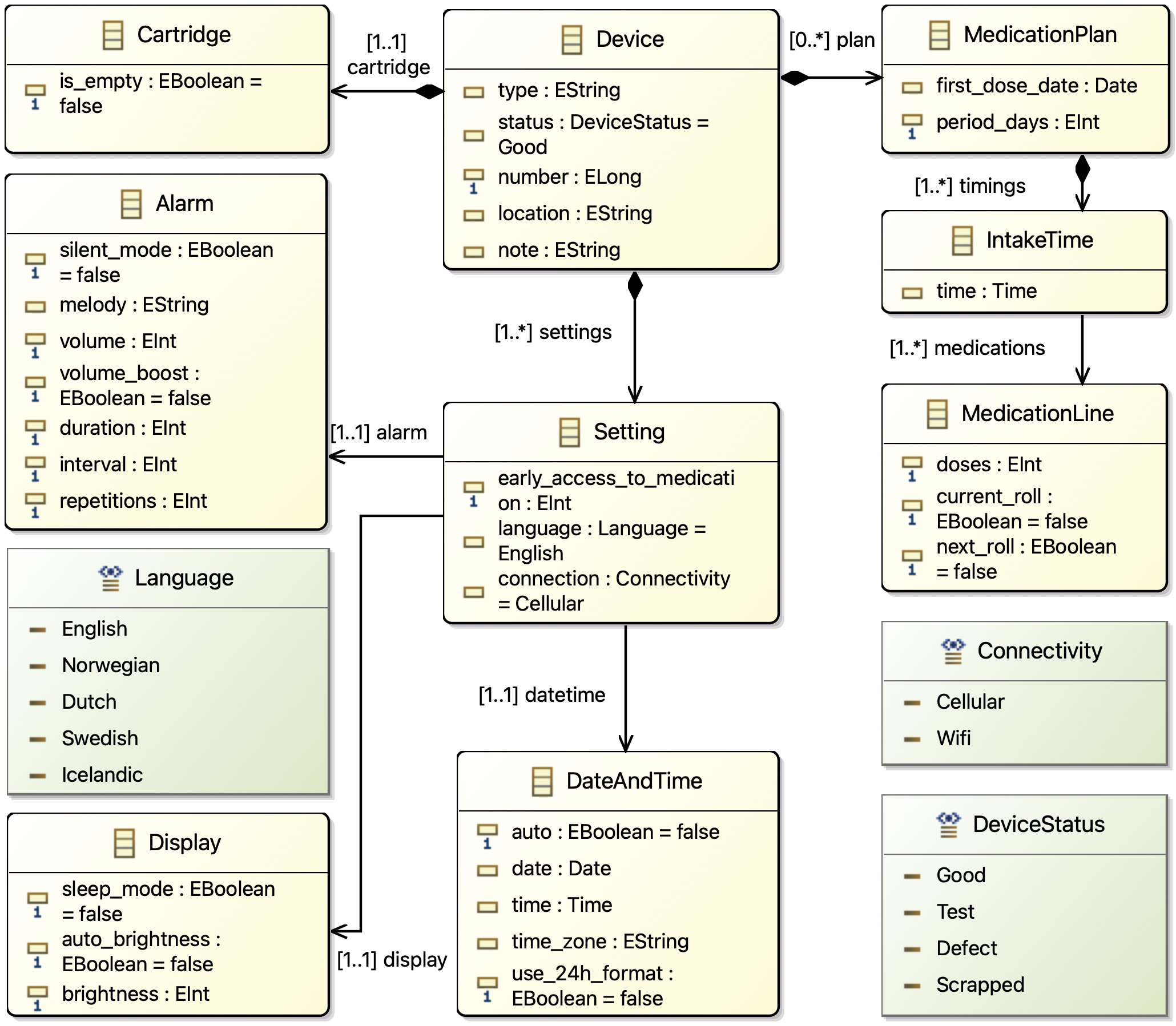}}
\caption{A domain model for smart medicine dispenser devices}
\label{fig:mm}
\end{figure}

\begin{figure*}[!t]
\centerline{\includegraphics[width=\linewidth, height=\textheight, keepaspectratio]{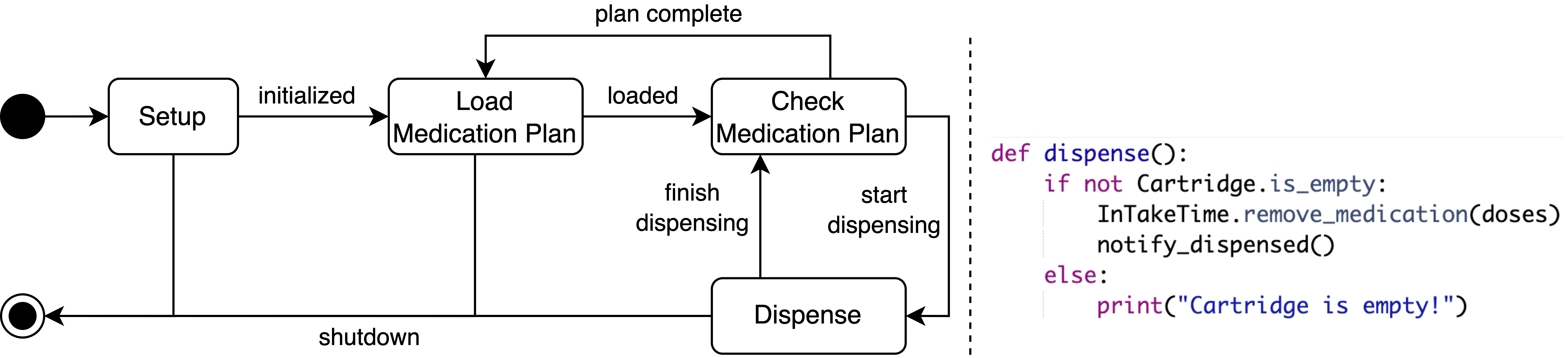}}
\caption{(Left) A state machine of a medicine dispenser - (Right) A code snippet for state \textit{Dispense}}
\label{fig:sm}
\end{figure*}

\subsection{Constructing DT Structure}\label{structural aspect of DT}
To make a DT structurally similar to a physical device, we develop a domain model capturing the structural aspects of medicine dispenser devices. 
Using the domain model, our approach creates an instance model representing the structural aspects of a particular device. 
In the following, we discuss these steps further. 

\subsubsection{Domain Model of Medicine Dispensers}\label{metamodel}
Figure~\ref{fig:mm} shows a domain model capturing abstract concepts of smart medicine dispenser devices. We identified these concepts from the documentation of multiple medicine dispensers provided by Oslo City's healthcare department. 
The central concept of the domain model is \textit{Device}, which has properties such as \textit{type}, \textit{status}, a unique \textit{number}, \textit{location}, and \textit{note}.  
The device status can be of four types represented with enumeration (\textit{DeviceStatus}) of four literals, i.e., \textit{Good}, \textit{Test}, \textit{Defect}, and \textit{Scrapped}. 
A device comprises a \textit{Cartridge} holding medicines, which has Boolean property \textit{is\_empty} denoting whether medicines are available or not. 
A device may store one or more \textit{MedicinePlan}s, which can be obtained from the healthcare provider's cloud application. A medicine plan has an initial date (\textit{first\_dose\_date}), a number of days of a period (\textit{period\_days}), and multiple-dose intake timings (\textit{IntakeTime}). 
For example, the medicine dose intake time can be 09:00, 13:00, and 19:00 daily.  
Each intake time can have a specified number of medicine doses represented as \textit{MedicineLine}. 
The number of doses can be either specified in the current medicine roll (\textit{current\_roll}) or the next medicine roll (\textit{next\_roll}) or in both rolls. 

A device can have one or more different \textit{Settings}, which can be configured by a patient, including setting up time (minutes) to take medicine early (\textit{early\_access\_to\_medication}), display language, and connection type (\textit{Cellular} or \textit{Wifi}). 
Currently, medicine dispensers support five different languages represented as enumeration \textit{Language} with literals such as \textit{English} and \textit{Norwegian}. 
A device setting also includes the setup of \textit{DateAndTime}, \textit{Display} (characterized with \textit{brightness}, \textit{sleep\_mode}, etc.), and \textit{Alarm} (characterized with \textit{silent\_mode}, \textit{melody}, \textit{repetitions}, etc.). 

The device domain model is accompanied by constraints specified on different concepts and properties in the Object Constraint Language (OCL).
Listing~\ref{lst:constraints} shows three of them for illustration purposes:
Constraint \textit{C1} specifies the possible range of values for \textit{period\_days} of \textit{MedicationPlan}; constraint \textit{C2} limits the number of allowed medicine doses corresponding to \textit{MedicationLine}; and constraint \textit{C3} restricts the setting for early access to a medication duration. These constraints are derived based on our domain knowledge and validated with the Web application (our industrial healthcare IoT application) of the IoT system.
We use the Eclipse Modeling tool and Eclipse Modeling Framework (EMF) for developing the domain model and specifying OCL constraints.

\begin{lstlisting}[label=lst:constraints, language=ocl, caption={Selected constraints on device properties}, linewidth=15cm, numbers=none]
-- range for the number of days of a medication plan
C1: context MedicationPlan inv: self.period_days >= 1 and 
                                        self.period_days <= 28
-- range of allowed medicine doses
C2: context MedicationLine inv: self.doses >= 0 and self.doses <= 9
-- range for early access to medication 
C3: context Setting inv: self.early_access_to_medication >= 1 and 
                                self.early_access_to_medication <= 300
\end{lstlisting}

\subsubsection{Generating Instance Model}\label{instance model generation}
The first step of our approach is model-to-text transformation, i.e., serialization of a domain model into a textural format. 
Using the device domain model (as shown in Figure~\ref{fig:mm}), our approach automatically generates an input template to conveniently capture device properties. 
The generated input template is in the format of JavaScript Object Notation (JSON), such that test engineers can easily use it to provide information about a particular device. 
Such information, including the number of instances of each domain class in the domain model to be created and the values of class properties, is required for instantiating the domain model.
The transformation process starts with the root domain class to create a root object in JSON along with its domain properties. 
For each relationship of the root domain class, a nested JSON object is created, including domain properties. 
In the case of default values or enumerations, domain properties are assigned with the corresponding property values that can be modified by test engineers. 

The next step is model-to-text transformation, i.e., the deserialization of JSON format into an instance model. 
With the provided information in the JSON input, our approach generates an instance model of the device domain model. 
For this purpose, our approach starts with the root class in JSON input and creates an instance of the corresponding domain class. 
Next, the properties of the domain class are converted into slots in the instance, and the values of these slots are populated from the JSON input. 
After instantiating the root domain class, we identify its associations with other domain classes and their multiplicities. 
For each associated domain class, we create one or multiple instances based on the number of instances given in JSON input and the multiplicity values. 
Slots and their values are populated for each property of the domain class in the same way it is done for the root domain class. 
This process continues until all information in the JSON input is utilized in the form of an instance model. 
Such an instance model represents the structure of a medicine dispenser device. 
During the model instantiation process, each property value is validated with relevant OCL constraints to ensure that an instance model contains valid device configurations. To do this, we have built a customized constraint validator compliant with the instance generation process. 
This involves converting OCL constraints into Python conditions and using a Python evaluator for validating instances' property values.
For model instantiation, we use Python-based EMF library named PyEcore~\cite{pyecore}.

\subsection{Constructing DT Behavior}\label{behavioral aspect of DT}

Our approach models device behaviors as an executable UML state machine to make a DT executable like its corresponding physical medicine dispenser. 
For this purpose, our approach requires a UML state machine modeling abstract behavior and a set of code snippets corresponding to each state, a common practice of making executable models~\cite{leduc2017revisiting}. 

As the first step, a test engineer can use UML or Ecore (e.g., in~\cite{leduc2017revisiting}) tools to create a state machine capturing abstract states and transitions, which models the abstract behavior of a physical medicine dispenser. The next step is to add code snippets in Python corresponding to each state of the state machine. For instance, the left-hand side of Figure~\ref{fig:sm} presents a state machine representing the abstract behavior of a medicine dispenser. The medicine dispenser starts by setting itself up with previous configurations (state \textit{Setup}). After that, it transits to state \textit{Load Medication Plan}, i.e., fetching a medication plan from a healthcare IoT application. If no medication plan is available, the dispenser periodically checks for an update until a medication plan is loaded; otherwise, the dispenser transits to state \textit{Check Medication Plan}, during which the loaded medicine plan is checked, and the dispenser waits for the medicine intake time. 
When the current date and time match the intake time, the dispenser starts the medicine dispensing process (i.e., state \textit{Dispense}). 
On the right-hand side of Figure~\ref{fig:sm}, we show a code snippet for state \textit{Dispense}, which checks whether the dispenser cartridge is empty. If it is not empty, the dispenser removes medication with specified doses from the medication intake time and notifies when the process is finished. 
When the dispensing task is finished, the dispenser returns to state \textit{Check Medication Plan}. If the medication plan is completed, the dispenser waits for a new medication plan to load. At any state, the dispenser can be shut down. 

Though test engineers must build such state machines manually, the required modeling effort is one-time~\cite{sartajtesting, sartaj2021automated}. Moreover, such a state machine can be reused for various medicine dispensers from the same or different vendors. Furthermore, state machines for devices such as medicine dispensers are not challenging to build as their behaviors are typically not complex. 

Our approach loads the input state machine and creates an internal representation. Specifically, we use code snippets for each state and create incoming/outgoing transitions by making function calls among code snippets of different states. 
This results in an executable behavioral model of the DT. To link it to the structural model of the DT, our approach needs to add the executable behavior as the owned behavior of the root domain class instance (i.e., \textit{Device}). 
Consequently, we obtain \textit{Device Executable Model}, which is a DT of a physical medicine dispenser device (Figure~\ref{fig:app}).
For modeling and executing DT behavior, we use Action Language for EMF (ALE~\cite{leduc2017revisiting}) with the Eclipse Modeling tool and PyEcore.

\subsection{Operating DT}\label{operating DT}

To simulate the device operation, our approach first creates APIs to support communication and integration of the DT with the SUT. 
Next, using APIs, our approach creates a DT communication server to handle all the communications of the DT with the SUT and physical devices, as shown in Figure \ref{fig:iot}. The physical device, DT, and SUT are all integrated via respective APIs.

\subsubsection{Generating APIs}\label{API Generation}
\begin{figure}[htbp]
\centerline{\includegraphics[width=13cm, height=3.9cm, keepaspectratio]{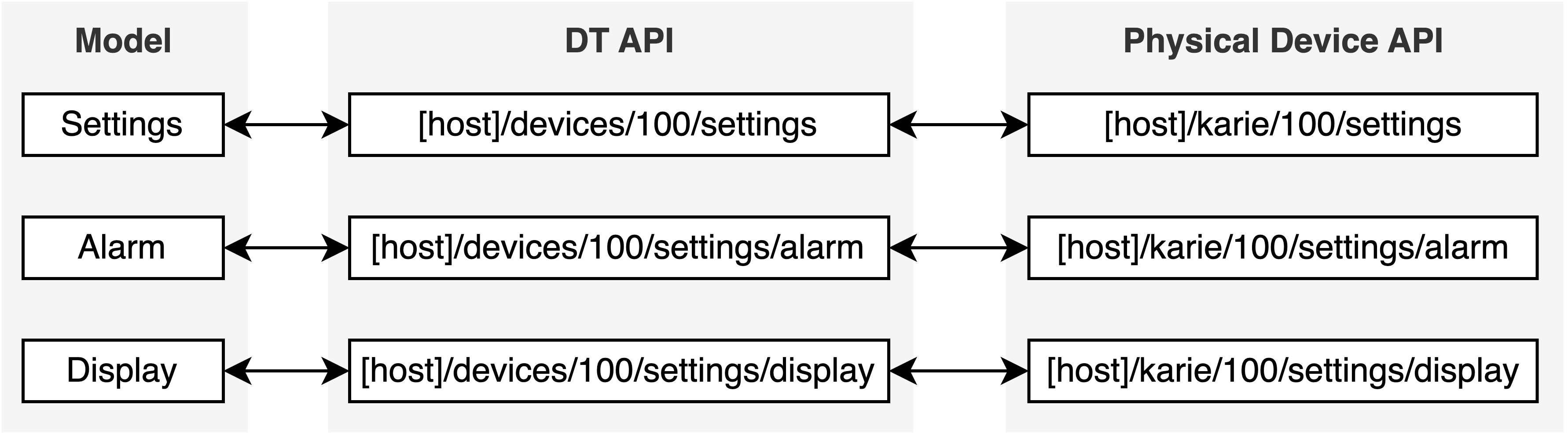}}
\caption{An example of Model-DT-Device mapping}
\label{fig:dtmap}
\end{figure}

The DT APIs are generated with \textit{APIs Generation} of our approach (Figure~\ref{fig:app}) via a customizable and pre-defined mapping, i.e., \textit{Mapping (Model-DT-Device)}. 
These DT APIs must be mapped to the domain model's classes, which helps a DT to identify a particular request for a specific model instance. 
Figure~\ref{fig:dtmap} shows an example of mapping among model classes (i.e., Settings, Alarm, and Display), DT APIs for device number 100, and physical device APIs (e.g., Karie device \# 100). 
For illustration, suppose the SUT makes an API call (via a DT API) to change the Alarm settings of the DT. 
In that case, this request is received via the DT API (i.e., \url{[host]/devices/100/settings/alarm}) corresponding to \textit{Alarm}, a domain class of the domain model (Figure \ref{fig:mm}). 
To handle such a request, the DT updates the property values of the \textit{Alarm} instance with the data received as a part of the request. 
For the DT's communication with a physical device, the APIs of a particular device are required (e.g., for Karie as shown in Figure~\ref{fig:dtmap}). 
In the case of the Alarm settings example, the DT will communicate through Karie device API (e.g., \url{[host]/karie/100/settings/alarm}) to change the alarm settings on the device.

With the mapping, first, our approach creates APIs for a specific DT. 
This includes creating APIs and API endpoints for each association of a domain class.
For example, for Alarm settings, our approach creates an API for domain class \textit{Setting} and another endpoint for domain class \textit{Alarm}. 
Next, our approach links the DT APIs and the physical device APIs. 
These links are also created based on the domain classes. Also, taking the Alarm settings as an example, the API endpoint of the DT is linked to the corresponding API endpoint of the physical device. 
The resultant mapping is then utilized by \textit{DT Communication Server} to establish communications among the SUT, DTs, and physical devices.

\subsubsection{Enabling Communication} 
To handle DT's communication with the SUT and physical devices, our approach creates a DT communication server based on configurations (containing server addresses) provided by \textit{Test Engineer} and the DT APIs for connecting to physical devices, as shown in Figure \ref{fig:iot}. 

The SUT identifies a physical device via a unique serial number and communicates with it via device APIs. Similarly, to enable communication between the DT with the SUT, our approach also relies on the device's serial number to locate the corresponding DT APIs, which are specific to the device of concern. Essentially, the DT APIs replace the device APIs when communicating with the SUT. Request and response methods of the APIs are also utilized, like for the physical device. When everything is ready, the DT communicates with the SUT and optimally with the physical devices if needed.

\textit{DT Communication Server} consists of various HTTP request handling methods such as GET, POST, PUT, and DELETE. It receives a request (e.g., get display settings) from the SUT, and based on the request type, it communicates with the DT. Specifically, to handle GET requests from the SUT, the server obtains the device information (e.g., brightness level, auto-brightness, and sleep mode) from the running DT and sends the response back containing the required data. If the target data is unavailable, the DT sends an error response. A POST/PUT request aims to modify device configurations (e.g., update display settings with brightness level to a maximum of five), for which the server sends data from the request body to the DT for updating its configurations. 
The DT validates input data using constraints specified along with the domain model (Section~\ref{metamodel}). If it is valid, the DT updates its properties according to the data; otherwise, the DT sends an error response back to \textit{DT Communication Server}. The DELETE request is only required when an existing medication plan needs to be removed. Upon receiving a DELETE request, \textit{DT Communication Server} first checks whether the medication plan exists. If yes, it is removed from the specified medication plan and sends the success response; otherwise, the DT returns an error response. 
Each response returned by \textit{DT Communication Server} consists of response time and status code (success/error code).

\subsubsection{Synchronizing DTs with Physical Devices}
To synchronize the DT with a physical device, our approach uses its execution logs, based on which our approach identifies the lower and upper bounds of the execution time of different operations (e.g., dispensing) and then calculates the average execution time required for the physical device to complete a particular operation. For example, \textit{dispensing} operation involves checking the cartridge, taking medicines from the cartridge, and removing medicine to dispense. For each operation, our approach adds delay (based on the average execution time of the physical device) to synchronize the DT with the physical device. 

We can also create multiple DTs representing multiple physical devices from the same or different vendors. To do so, our approach requires serial numbers for each DT, which need to be provided as input by test engineers (Figure~\ref{fig:app}). Second, our approach uses serial numbers to create a specified number (\textit{N}) of uniquely identifiable DTs (Figure~\ref{fig:app}) and their APIs, which subsequently are used to integrate multiple DTs with the SUT. Also, the DT communication server uses these APIs to handle requests for a particular DT.

\section{Evaluation}\label{evaluation}
The overall focus of our approach is to support automated system-level testing of healthcare IoT applications with multiple DTs in place of physical devices. 
Analyzing the fidelity of a DT's operation and scalability in creating multiple DTs are important concerns of our industry partner. 
Thus, in this evaluation, we aim to assess the fidelity of DTs from two aspects: their functional similarity with physical counterparts and the feasibility of simulating many devices with DTs in parallel. 
We analyze the fidelity of DTs in terms of response time and status code because these are the key factors for testing purposes. 
These factors help in observing whether DTs are operating similarly to physical devices or not. 
Response time determines the total time a DT or a physical device takes to process a request. 
Status code specifies the processing mechanism and output (success/error) generated by a DT or a physical device to handle a request. 
We formulate the following three research questions (RQs) based on our evaluation objectives.

\begin{itemize}[leftmargin=10pt]
    \item \textbf{RQ1:} \textit{What is the similarity between the DT and the physical device regarding the response time?}\\
    The purpose of this RQ is to analyze the synchronization of a DT's response time with the physical device's response time. 
    \item \textbf{RQ2:} \textit{What is the similarity between the DT and the physical device concerning the status codes?} \\
    In this RQ, we evaluate the output in the form of status codes generated by a DT and a physical device to observe functional resemblance. 
    \item \textbf{RQ3:} \textit{How does similarity (regarding both the response time and status code) vary with the increase in the number of DTs in operation?}\\
    This RQ aims to assess the scalability of our approach in utilizing multiple DTs for testing. 
\end{itemize}

\subsection{Implementation}
We implemented our approach in Python with two modules. The first module loads the device domain model, generates the JSON input template, and validates the constraints upon receiving input data from the SUT. For all the activities related to the domain model and generated instance model, we rely on PyEcore~\cite{pyecore}, a Python version of Eclipse Modeling Framework, because it provides advanced features (e.g., dynamic creating executable models efficiently), with which a \textit{Device Executable Model} (Figure \ref{fig:app}) (i.e., modeling the DT behavior in an executable state machine) is developed. The second module loads the model and initiates \textit{DT Communication Server}, which was developed with the Python-based framework, Flask~\cite{flask}. 
For API creation, we use Flask-RESTful, which is compatible with the Flask framework. 
The DT APIs follow the JSON format to be consistent with the SUT and the physical device for information interchanges. 

\subsection{Industrial Case Study}
Our industrial case study is provided by Oslo City's healthcare department. The SUT is a healthcare IoT application (details in Section \ref{subsec:industrialcontext}). 
We used Karie~\cite{karie} as the physical device, which is a widely-used multi-feature smart medicine dispenser. 
Karie can be integrated with various healthcare IoT applications through APIs. 
Karie loads the medication plan for the concerned patient, follows the plan for dispensing doses at a specified time, and notifies the healthcare IoT application about events such as missed doses. 
Besides medicine dispensing, Karie lets users personalize settings such as time zone, alarm duration, and volume. These settings can be done through the device interface or the user interface of the healthcare IoT application.

\begin{table*}[!t]
	\centering
	\noindent
	\caption{An association of RQs with fidelity aspect, metrics, and statistical test}
	\begin{tabular}{l p{2.7cm} p{3.2cm} p{2.7cm} p{2.7cm}}\toprule
		\multicolumn{1}{ l }{\textbf{RQ}} & \textbf{Fidelity Aspect} & \textbf{Comparison} & \textbf{Metrics} & \textbf{Statistical Test} \\ 
		\cmidrule(lr){1-1}\cmidrule(lr){2-2}\cmidrule(ll){3-3}\cmidrule(ll){4-4}\cmidrule(ll){5-5}
		\multicolumn{1}{ l }{1}& Response time &One PD with one DT& \% Similarity, Mean, STD & Wilcoxon signed-rank test\\
		\multicolumn{1}{ l }{2} & Status code &One PD with one DT& \% Similarity, Mean, STD & Fisher's exact test \\
            \multicolumn{1}{ l }{3} &  Response  time \& Status code &One PD with 10, 20, 30, ..., 100 DTs& \% Similarity & - \\
		\bottomrule
	\end{tabular}
	\label{tab:setup}
\end{table*}

\subsection{Evaluation Setup, Metric, and Execution}

\textbf{Setup.}
We developed a random testing strategy to generate test data for testing a healthcare IoT application, which randomly chooses values for each device property. Note that developing a cost-effective testing strategy is not the focus of this paper; however, 
in the future, rigorous test methods can be easily integrated with our approach. 

For RQ1 and RQ2, test data generated at a particular time is forked and transformed into two requests: one request is sent to the Karie DT, and the other is sent to the physical Karie device. 
The physical Karie and its DT process each incoming request independently and generate responses that are compared to know the similarities regarding response time and status codes. Response time indicates the time it takes for the physical Karie or its DT to process a request, while the status code indicates the request handling functionality of the device. In case of success, the physical device (PD) and DT return a status code belonging to the 2XX category. In the case of an error in handling the request, a server error-related code representing 5XX is returned. For instance, an error code 503 is returned if a POST request contains data with invalid values that cannot be assigned to a PD or a DT. 
Typically, the error code 503 is related to unavailable services, i.e., a physical device takes more time to process a request. 
For example, Karie's medicine dispensing state usually takes more than one minute to dispense medicines, and during this state, the device cannot receive/process another request. 
This results in a 503 error code indicating that a request cannot be handled at the moment.

We conducted the experiment in multiple runs, each with a different time duration (i.e., one, two, four, six, eight, and ten hours), and we kept control of the number of requests. 
Since our experiment involves a physical device (Kaire), we set up our experiment following the guidelines of our industry partner. 
For up to four hours of execution runs, we restrict the number of requests to 30 calls per minute. 
For more prolonged execution runs (i.e., six, eight, and ten hours), we limit the number of requests to 20 calls per minute. 
The maximum long execution is ten hours because the maximum number of allowed requests exceeds after that. 
Each call is a POST request with parameter values generated with our random testing strategy. 

For RQ3, we create 100 DTs in multiple batches, i.e., 10, 20, 30, ..., 100 DTs. 
We use previous test data from the one-hour run, create requests for all DTs in each batch, and execute them concurrently in a thread pool. We receive a response containing the response time and status code for each request made to a DT.
We conducted the experiment on a machine with a macOS operating system, an 8-core CPU, and 24 GB RAM.

\textbf{Metrics and Statistical Tests.}
Table~\ref{tab:setup} shows the fidelity aspects considered in each RQ and the metrics used to evaluate RQs. 
To measure the fidelity of the DT, we use the similarity metric proposed in~\cite{munoz2022using}, which is developed based on the Needleman-Wunsch algorithm~\cite{needleman1970general}. 
This metric was initially developed to assess the similarity between traces of a DT and a physical twin. 
In our case, the traces of the DT and physical device contain information related to response time and status codes at various states. 
This is quite a similar context to the initial objective of the metric. 
This metric has also been used for measuring DT fidelity~\cite{munoz2022measuring}. 
Therefore, we adapt this metric in our context to measure DT fidelity. 
Regarding the response time, we set the tolerance level of similarity to one second, implying that the maximum amount of difference between the DT and PD is still considered ``similar" by the metric. 
Note that response status codes are categorical; therefore, there is no need to specify a tolerance level. 
For RQ1 and RQ2, we also statistically analyze the similarities between DT and PD to test the hypothesis that the DT's functionality is different from the PD regarding response time and status code with the Wilcoxon signed-rank test and Fisher's exact test, respectively, with the significance level ($\alpha$) 0.05.

\begin{table*}[!t]
	\centering
	\noindent
	\caption{Results of comparison between DT and PD for RQ1 and RQ2}
	\begin{tabular}{l N N N N}\toprule
		\multicolumn{1}{l }{\textbf{}} & \multicolumn{2}{c }{\textbf{RQ1: Response Time}} & \multicolumn{2}{c }{\textbf{RQ2: Status Code}} \\ 
		\cmidrule(lr){2-3}
		\cmidrule(ll){4-5}
		\multicolumn{1}{ l }{\textbf{Run}} & \textbf{Similarity} & \textbf{Wilcoxon Test (p-value)}& \textbf{Similarity} & \textbf{Fisher Exact Test (p-value)}\\ 
		\cmidrule(lr){1-1}
		\cmidrule(lr){2-3}
		\cmidrule(ll){4-5}
		\multicolumn{1}{ l }{\textbf{One hour}} &95.09\%&0.99 & 92.38\%&0.57\\
		\multicolumn{1}{ l }{\textbf{Two hours}}  & 91.92\%&0.99 & 91.02\%&0.88\\
		\multicolumn{1}{ l }{\textbf{Four hours}}  & 93.25\%&1.0 & 92.29\%&0.57\\
		\multicolumn{1}{ l }{\textbf{Six hours}}  & 91.67\%&0.99 & 90.02\%&0.11\\
		\multicolumn{1}{ l }{\textbf{Eight hours}}  & 92.47\%&1.0 & 92.28\%&0.92\\
		\multicolumn{1}{ l }{\textbf{Ten hours}}  & 93.03\%&1.0& 92.17\%&0.99\\
		\cmidrule(ll){1-5}
		\multicolumn{1}{ l }{\textbf{Mean}} & 92.90 & - & 91.69 & -\\ 
            \multicolumn{1}{ l }{\textbf{STD}} & 1.12 & - & 0.89 & -\\ 
		\bottomrule
	\end{tabular}
	\label{rq12results}
\end{table*}

\subsection{Results}\label{results}
Following, we discuss the results and their analysis corresponding to each research question.

\subsubsection{RQ1 - Fidelity regarding Response Time Similarity}\label{RQ1}
Table~\ref{rq12results} shows that the similarity values between DT and PD for all runs are more than 91\%. 
Overall, the mean similarity between DT and PD is $\approx$92\%. 
The standard deviation value suggests the percentage similarity is centered around mean, i..e, $\approx$92\%.
Furthermore, the results of the Wilcoxon signed-rank test (p-value$>\alpha$) indicate that DT and PD perform statistically similarly in terms of response time. 

\begin{figure}[htbp]
\centerline{\includegraphics[width=8.6cm, height=6.1cm, keepaspectratio]{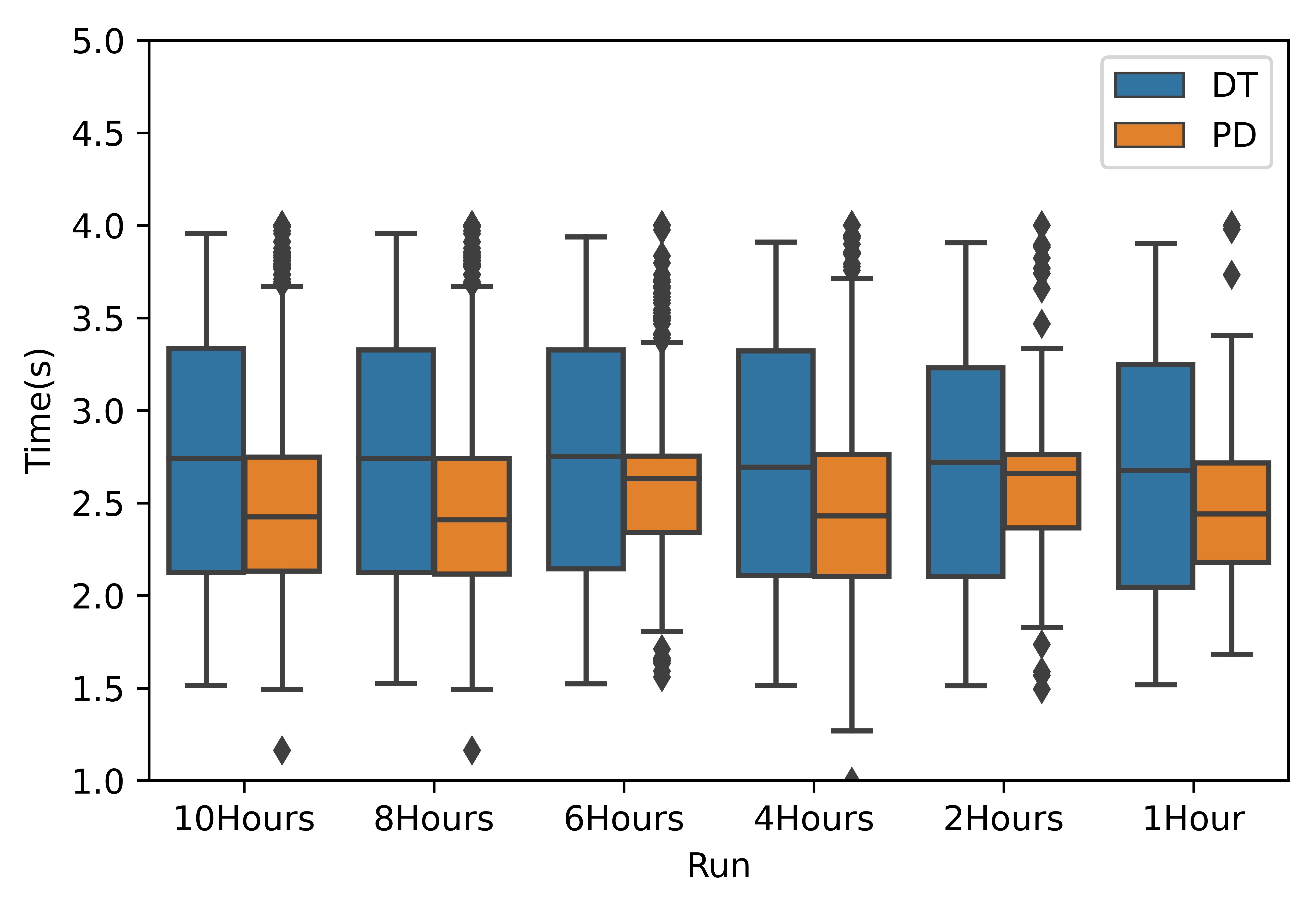}}
\caption{RQ1: Comparison between DT and PD based on response time}
\label{fig:rq1time}
\end{figure}

Figure~\ref{fig:rq1time} compares PD and DT based on the response time's absolute values (in seconds). From the figure, we see that the median response time of DT is within the range of 2.5--3 seconds for all hour runs, which is consistent with PD for its two-hour and six-hour runs. 
For the other hour runs, it is $\approx$2.4 seconds. 
Furthermore, we observe that the minimum and maximum values of DT and PD are nearly the same.  

The results show that the similarity in response time does not fluctuate largely with the increase in the time durations of the runs. 
Sometimes, the PD takes a little more time to return the response because the DT server runs locally, whereas the device communicates with the healthcare IoT application via the PD web server over a network. Consequently, network delays may delay the response time. 
Furthermore, in the case of response status codes, all status codes of DT are not 100\% aligned with the PD.

\begin{tcolorbox}[beamer,colframe=black!50!black,title=RQ1: Response Time]
    In general, the similarity between DT and PD is more than 92\%, regarding response time, indicating that they are almost aligned based on the request processing time.
\end{tcolorbox}

\subsubsection{RQ2 - Fidelity in terms of Status Code Similarity}\label{RQ2}
The results of the comparison between DT and PD based on status codes are presented in Table~\ref{rq12results}, which show that DT and PD are more than 91\% comparable in most of the hour runs. In one case, i.e., six-hour run, the similarity is $\approx$90\%. The mean similarity value indicates that DT and PD are more than $\approx$92\% functionally equivalent. The standard deviation value shows that most similarity values are concentrated around $\approx$92\%. 
The results of Fisher's Exact test (p-value$>\alpha$) suggest that DT performs significantly similarly to PD in terms of status codes. 

\begin{figure}[htbp]
\centerline{\includegraphics[width=8.6cm, height=6cm, keepaspectratio]{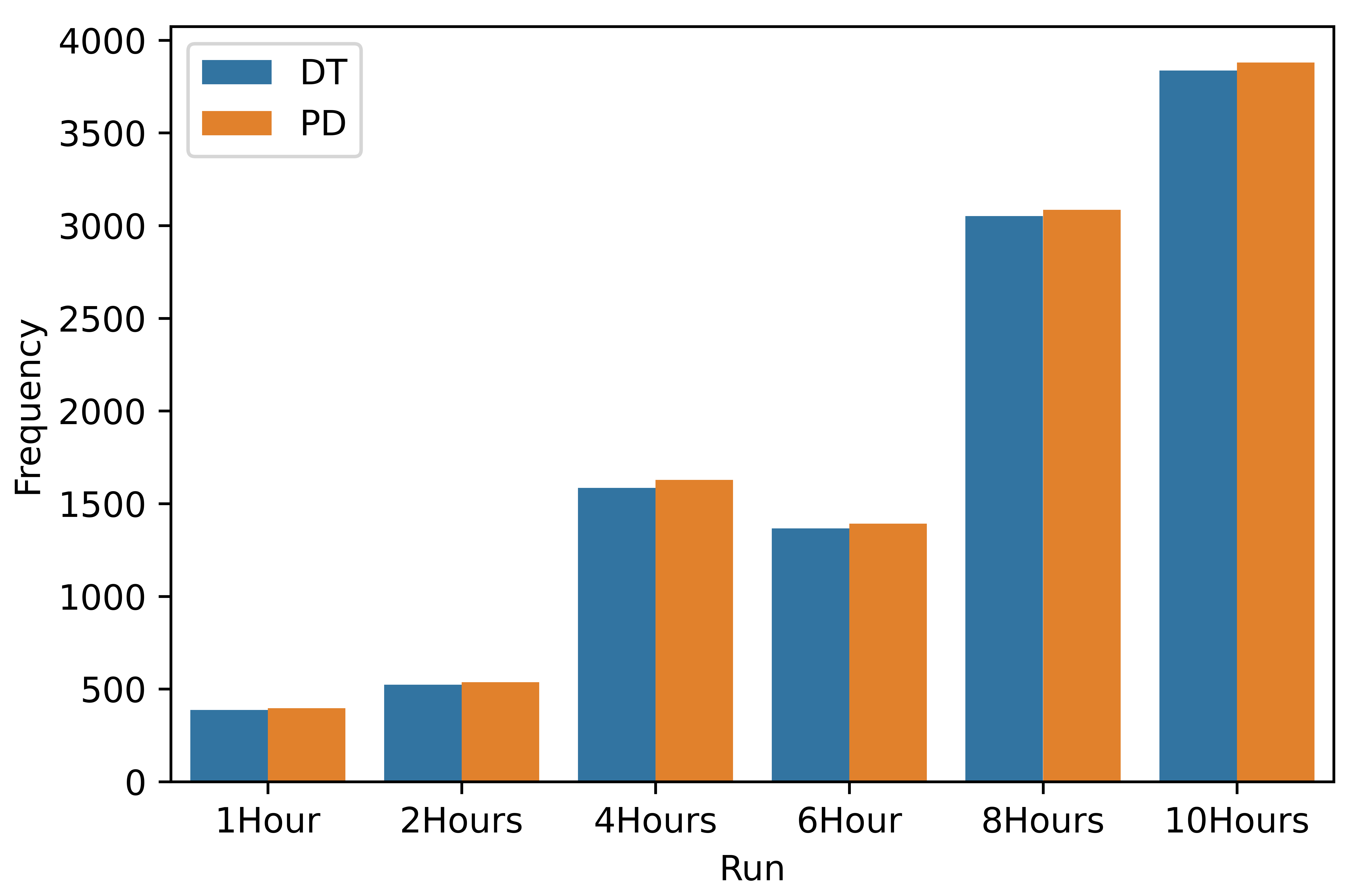}}
\caption{RQ2: Comparison between DT and PD based on the success status code 200}
\label{fig:rq2sc1}
\end{figure}

\begin{figure}[htbp]
\centerline{\includegraphics[width=8.6cm, height=6cm, keepaspectratio]{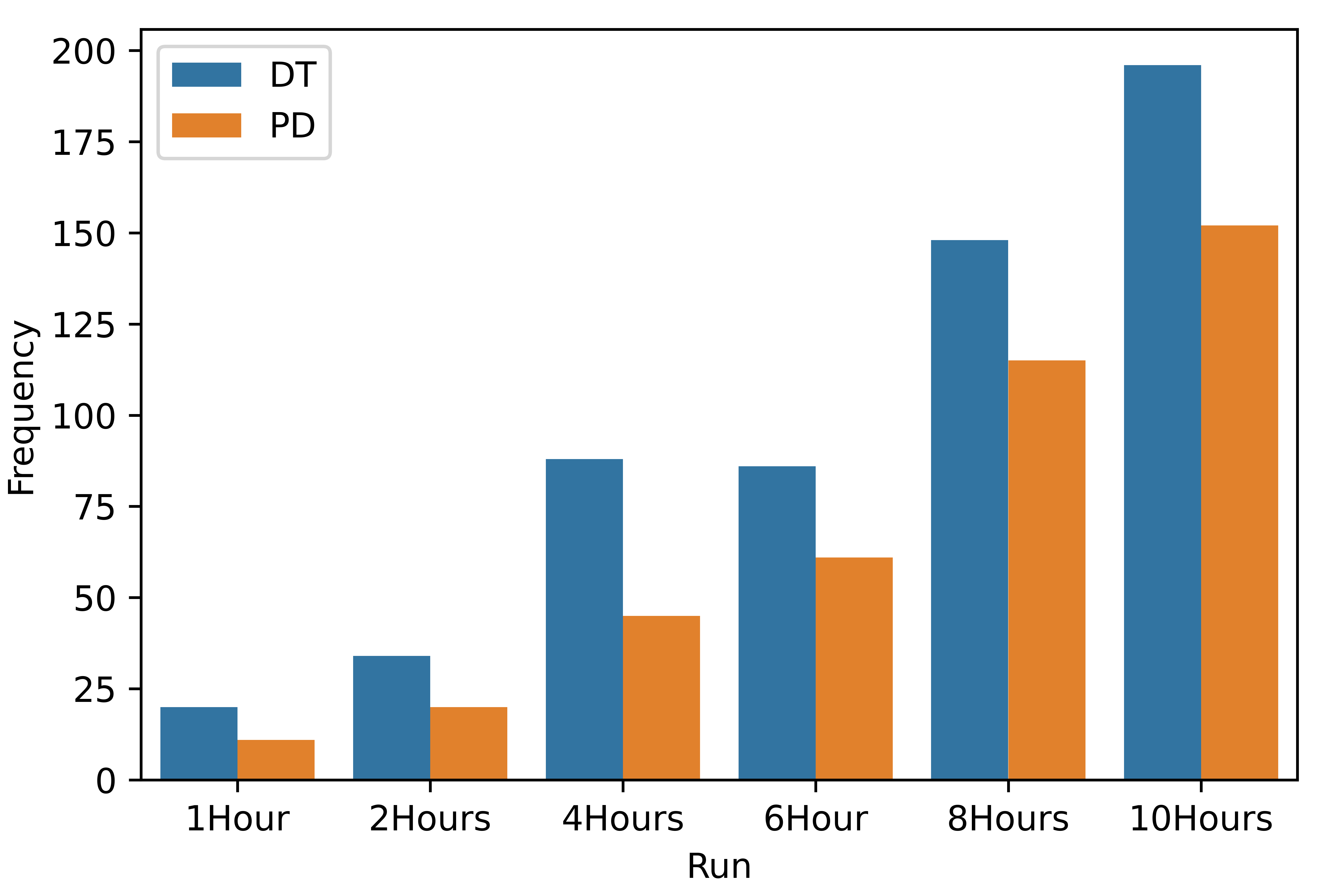}}
\caption{RQ2: Comparison between DT and PD based on the error status code of 503}
\label{fig:rq2sc2}
\end{figure}

Figure~\ref{fig:rq2sc1} and Figure~\ref{fig:rq2sc2} show a PD and DT comparison based on the success and error status codes. 
It can be seen from the success codes (200) of PD and DT that both function similarly in all runs. 
The number of error codes (503) DT generates varies slightly from PD. 
The analysis of some cases reveals that the DT adheres to all constraints (following the documentation of the PD). 
However, sometimes, the PD partially accepts some correct configurations while declining the out-of-bound values. 
For example, if the display configurations are correct and volume settings are incorrect, the device sometimes accepts the display configurations but ignores the incorrect volume settings. 
Further analysis is required to understand this observation better. 

\begin{tcolorbox}[beamer,colframe=black!50!black,title=RQ2: Status Codes]
    The overall similarity between DT and PD based on the status codes is more than 92\%, which indicates that DT performs nearly equivalent to its corresponding PD. 
\end{tcolorbox}

\subsubsection{RQ3 - Fidelity in terms of Response Time and Status Code of Concurrently Operating DTs}
\begin{figure}[htbp]
\centerline{\includegraphics[width=8.8cm, height=6.1cm, keepaspectratio]{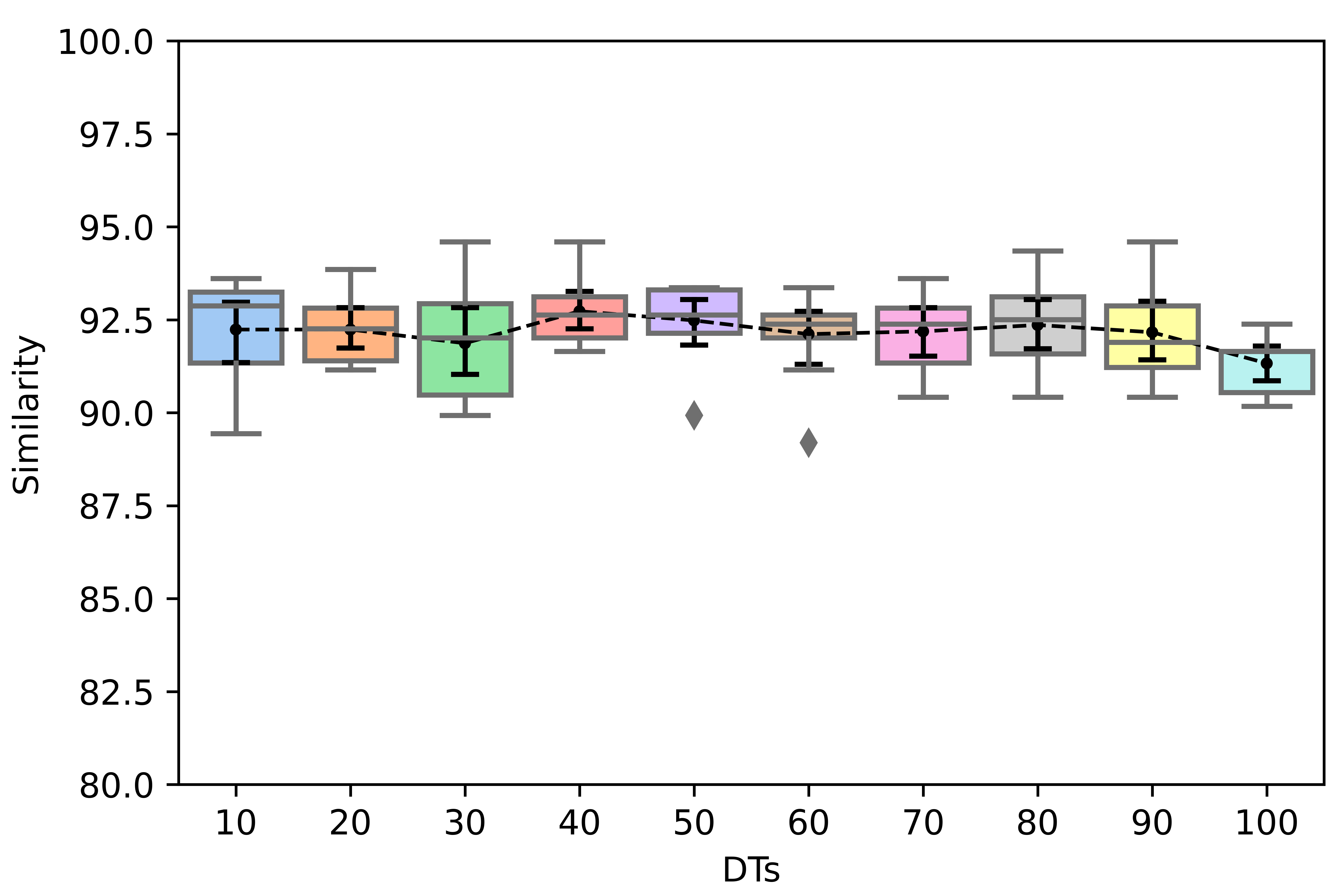}}
\caption{RQ3: Comparison of PD with 100 concurrently operating DTs based on the response time}
\label{fig:dtstime}
\end{figure}

Figure~\ref{fig:dtstime} shows the response time-based comparison of PD with 100 DTs in different batches, i.e., 10, 20, 30, ..., 100 DTs. 
The boxplots show that the similarity values of PD's comparison with different batches of DTs vary slightly, with most values around 92\%.  
This suggests that the PD's similarity in terms of response time is consistent across 100 DTs. 
The results of the status code-based comparison of PD with 100 DTs show a similarity value of $\approx$92\%, suggesting that PD's functionality in terms of status code is comparable with the increase in the number of DTs. 
The outcomes of PD's comparison with 100 DTs based on response time and status code are consistent with the results of RQ1 (Section~\ref{RQ1}) and RQ2 (Section~\ref{RQ2}). 

\begin{tcolorbox}[beamer,colframe=black!50!black,title=RQ3: Fidelity of Multiple DTs]
    The overall similarity of the PD with 100 DTs (of varying batch sizes) based on response time and status codes is $\approx$92\%, which indicates that the DTs behave very similarly to PD even when they are operating concurrently, i.e., simulating 100 PDs. 
\end{tcolorbox}

\subsection{Threats to Validity}

To reduce threats to \textbf{external validity}, we evaluate our approach with an industrial case study characterizing a widely-used smart medicine dispenser, which is a good representative. 
In the future, we plan to include more types of devices for large-scale evaluation. 
The key threat to \textbf{internal validity} is related to evaluation setups. 
We carefully designed the evaluation setup to minimize this threat. 
We use APIs and their documentation provided by our industry partner. 
We also held various sessions with practitioners of our industry partner to demonstrate the setup and get their feedback. 
Besides initial configurations, our approach requires no other parameter tuning during execution. 
To reduce the chances of \textbf{construct validity} threat, we used the similarity metric for DTs~\cite{munoz2022using}. 
We also used metrics such as mean and standard deviation to analyze the results.  
To handle the \textbf{conclusion validity} threat, we executed experiments in multiple runs of varying duration. 
We also used statistical tests such as the Wilcoxon signed-rank test and Fisher's exact test with the significance level suggested in~\cite{arcuri2011practical}.

\subsection{Data Availability}
We make an open-source implementation of our approach available along with example APIs. We, however, cannot provide a replication package for the industrial case study due to non-disclosure agreements.

\section{Discussions and Lessons Learned}\label{lessonslearned}
Based on the empirical evaluation using an industrial system, we provide detailed insights into the results and lessons learned in the following subsections. 

\subsection{Empowering Test Infrastructure with DTs}
Since the system-level testing of healthcare IoT systems requires integrating a large number of various physical devices, for each test execution, the SUT needs to communicate with them via a web server. Each device vendor has a dedicated web server that allows limited requests to support efficient and secure communications with their devices. Sending many requests to a web server during testing is practically infeasible, potentially leading to denial of service (DoS) attacks on the web server, resulting in blocking further requests. This puts a practical constraint on test case execution. Furthermore, a physical device might be damaged if it processes requests intensively during testing. Due to these observations from the operating system of Oslo City, we propose using DTs, as the replacement to physical devices for enabling testing without triggering the self-protection mechanism of the web server, employing a large number of physical devices and damaging them. Our vision is to build a test infrastructure for Oslo City empowered with DTs such that, in the long run, test effectiveness can be improved and cost can be reduced. 
Utilizing DTs in place of physical devices saves monetary and operating costs in creating test infrastructure for healthcare IoT applications.

\subsection{DT Modeling}
For modeling the structural aspects of a DT, our approach creates the domain model representation in JSON, which is convenient for test engineers to use to provide inputs on device properties because JSON is ECMA's Data Interchange Standard and is widely used in web applications~\cite{lin2012comparison}. To model DT behavior, test engineers need to know how to use a modeling tool (e.g., Papyrus) and how a medicine dispenser functions. The behavior of medicine dispensers can be easily interpreted by operating them, and they are often made simple such that users of special groups (e.g., elderly citizens) can use them without difficulties. Therefore, we consider that building a DT for a medical device does not require much modeling effort. 
In addition, state machines like the one in Figure~\ref{fig:sm} can already be used for Karie medicine dispensers and are a starting point to create/customize state machines for other medicine dispensers. Building a DT requires a one-time modeling effort and can be easily reused via configuration for building other DTs. 
This leads to time cost savings while creating DTs of different medicine dispensers.

\subsection{Cost Reduction with DTs}
DTs save considerable time by replacing physical medicine dispensers for testing healthcare IoT applications at scale, especially in rapid-release development phases. In our experience, we observed that integrating and configuring a medicine dispenser with a healthcare IoT application usually requires several manual steps including the steps of ensuring power supply/battery level, loading medicine rolls, initializing devices, ensuring network availability, connecting with a healthcare IoT application, and configuring device settings. 
Even assuming everything goes as planned, configuring 100 medicine dispensers takes a considerable amount of time. 
However, based on practical experience, many issues arise during configuration due to many inherent uncertainties in the operational environment (such as connection problems and continuous deployment of new software resulting in unpredictable behavior of healthcare IoT applications).  
One could argue that this is a one-time effort to establish a laboratory with hundreds of different devices. However, operating and maintaining such a laboratory incurs significant costs, including monetary costs, and requires specific expertise from test engineers. In contrast, our approach enables building a virtual lab with hundreds and thousands of medical devices whose operation and maintenance require little cost compared to such a physical laboratory. Quantitatively assessing the cost reduction of the approach requires a dedicated study, which, unfortunately, is impossible for us to conduct due to practical constraints (e.g., getting access to hundreds of medical devices and setting up a physical laboratory). 

Another important aspect to consider when assessing cost is the constant evolution of healthcare IoT applications. For instance, in the case of device upgrades, manual effort is always required to integrate and configure new or updated medicine dispensers. With our approach, creating models takes approximately 10-15 minutes. This is a one-time effort and can be reused for other types of medicine dispensers. To generate 100 DTs, our approach takes $\approx$0.65 seconds, which is negligible compared to the time cost required for integrating 100 physical medicine dispensers into IoT applications. The time cost saving with our approach is especially notable when running tests after each rapid release (daily/weekly), which is infeasible with physical devices. Furthermore, our approach eliminates practical constraints of device damage during testing, which is also monetary-wise valuable. Finally, we foresee the need for a dedicated study in the future with industry practitioners and test infrastructure to analyze the quantitative cost reduction.

\subsection{DT Fidelity}
Our experiments focused exclusively on assessing the fidelity of DTs, developed with our approach, by comparing with their physical counterparts, medicine dispensers integrated with our industrial healthcare IoT application (the cloud-based IoT healthcare system). We also experimented with up to 100 DTs of one type of medicine dispenser in a laboratory setting. The results showed that the DTs consistently behaved similarly to the medicine dispenser, even when all DTs operated concurrently. Though these results are promising, we need to assess the fidelity of our DTs even at a larger scale, as in practice, thousands of devices are integrated, and they might come from different vendors (e.g., alternatives to Karie could be Medido~\cite{medido} and Pilly~\cite{pilly}), have different types, and even the same type but different software versions (Section \ref{subsec:industrialcontext}). Moreover, the approach also needs to be extended for other types of devices that are used in Oslo City, such as GPS trackers and home monitoring. In general, our evaluation must consider the heterogeneity and scalability of integrating medical devices into healthcare IoT applications.

Our evaluation targeted fidelity assessment based on the functional similarity between DTs and a type of physical medicine dispensers (Karie). 
The functionality of DTs is important from a testing perspective because it is sufficient to perform system-level testing of healthcare IoT applications.  
Our fidelity assessment is focused on the functional similarity level due to limited access to device APIs and technical difficulties in obtaining physical device status at run time. We could not evaluate the fidelity of DTs in terms of their internal behaviors (to the level of the executable state machine). Upon overcoming these challenges, our next step is to assess and improve the DTs' fidelity in terms of their internal behaviors.

\subsection{Results Relevance and Approach Applicability}
The fundamental purpose of our approach is to provide a cost-effective and scalable solution for automated system-level testing of healthcare IoT applications using DTs in place of medicine dispensers
The results of our experiments indicate that DTs generated using our approach have more than 92\% functional similarity with the Karie medicine dispenser. 
In practice, we have observed that a physical medicine dispenser can also malfunction during testing~\cite{sartaj2023testing}. 
Since medicine dispensers (or other medical devices) are not the SUT, the DT fidelity level achieved by our approach is sufficient to support the automated testing of healthcare IoT applications (SUT). 
Furthermore, it is worth noting that the fidelity does not decrease with the increase in the number of DTs, which is a significant benefit for testing healthcare IoT applications with hundreds of medicine dispensers.

Our model-driven approach utilizes domain models and state machines as a basic step to create and operate DTs. 
The domain model (Figure~\ref{fig:mm}) and state machine (Figure~\ref{fig:sm}) are developed at an abstract level based on the specifications of multiple medicine dispensers. 
These models can be adapted for a variety of medicine dispensers such as Medido~\cite{medido} and Pilly~\cite{pilly}. 
In the case of medicine dispensers software upgrades, only the source code associated with a particular updated state needs to be fine-tuned. 
For medical devices other than medicine dispensers (such as pulse measuring devices), the domain model is required to be enriched with new concepts, and the behavioral model needs to be developed for the particular device. 
With the models ready, the remaining steps of the approach can be easily applied to create and operate DTs of the new medical device. 
This work contributes to the overall objective of creating a dependable, scalable, and cost-effective test infrastructure for healthcare IoT applications connected with a large number of heterogeneous medical devices~\cite{sartaj2023hita}.

\subsection{Developing Domain-specific Test Strategies}
The paper's current focus is assessing the fidelity of DTs. However, as discussed in Section \ref{sec:iot-hc}, the project's overall scope is to automate system testing of healthcare IoT applications. Currently, for our evaluation, we employ random testing. In the near future, as planned in this innovation project, we will develop novel testing strategies to find functional and performance faults in the IoT system cost-effectively. To this end, we will build domain-specific models, e.g., knowledge graphs related to medicine dispensers and other machines, and use them to guide testing. Given that we have access to REST APIs of the IoT applications, existing REST API-based test approaches, e.g., EvoMaster~\cite{arcuri2019restful}, RESTest~\cite{martin2021restest}, and RESTTESTGEN~\cite{viglianisi2020resttestgen}, are of our interest to investigate. We already see the opportunities of selecting a suitable tool and customizing it with domain-specific coverages (e.g., configuration coverage of medicine dispensers). These tools, in general, generate large numbers of tests. Therefore, building a test infrastructure empowered with our DTs is one of the essential steps to achieving automated testing of the healthcare IoT applications of Oslo City.

\begin{table*}[!h]
	\centering
    \small
	\noindent
    % \captionsetup{labelfont={color=blue},font={color=blue}}
	\caption{Comparison with related works}
    \begin{tabular}{p{0.01cm} p{1.8cm} p{0.9cm} p{1.3cm} p{1.2cm} p{1.3cm} p{1.6cm} p{2.7cm}}\toprule
          \multicolumn{2}{c }{\textbf{Paper}} & \multicolumn{1}{c }{\textbf{Approach}} & \multicolumn{1}{c }{\textbf{Tool}} & \multicolumn{1}{c }{\textbf{System}} & \multicolumn{1}{c }{\textbf{Domain}} & \multicolumn{1}{c }{\textbf{DT Category}} & \multicolumn{1}{c }{\textbf{Purpose}} \\ 
		\cmidrule(lr){1-2}\cmidrule(ll){3-3}\cmidrule(ll){4-4}\cmidrule(ll){5-5}\cmidrule(ll){6-6}\cmidrule(ll){7-7}\cmidrule(ll){8-8}
            \multirow{21}{*}{\textbf{\rotatebox[origin=c]{90}{\makecell{DT Generation and Operation}}}}
            &\textbf{~\cite{dalibor2022generating,michael2021towards,kirchhof2020model}}&MDE&MontiGem&CPS&SW Dev.&Low-code Platform&Support engineers
            % \\%\cmidrule(ll){2-4}
            \\%\cmidrule(ll){2-4}
            &\textbf{~\cite{corradini2023dtmn}}&DSL&ADOxx&IoT&General&General&Modeling % add-on
            \\%\cmidrule(ll){2-4}&&&&
            &\textbf{~\cite{yue2021understanding}}&MDE&N/A&CPS&Logistics&General&Synthesize DT concepts
            % \\%\cmidrule(ll){2-4}
            \\%\cmidrule(ll){2-4}
            &\textbf{~\cite{govindasamy2021air}}&MDE&DTDL&CPS/IoT&General&Air quality&Identify issues %Identify MDE DT problems
            % \\%\cmidrule(ll){2-4}
            \\%\cmidrule(ll){2-4}
            &\textbf{~\cite{dalibor2020towards,michael2022generating,bano2022process}}&MDE&MontiGem&CPS&Manufact-uring&Interactive GUI-based&Optimize manufacturing
            \\%\cmidrule(ll){2-4}
            &\textbf{~\cite{christofi2022novel}}&MDE&Nonpublic&CPS&Satellite&Spacecraft&Fault identification
            \\%\cmidrule(ll){2-4}
            &\textbf{~\cite{bonney2021digital}}&MDE&Cristallo&IoT&Web&Operational platform&Secure connectivity
            \\%\cmidrule(ll){2-4}
            &\textbf{~\cite{barat2022digital}}&MDE&Nonpublic&Socio-technical&Healthcare&Agent-based&Safe system analysis
            \\%\cmidrule(ll){2-4}
            &\textbf{~\cite{stary2022behavior}}&MDE&N/A&CPS&SW Dev.&Traffic control&Improve operations
            % \\%\cmidrule(ll){2-4}
            \\%\cmidrule(ll){2-4}
            &\textbf{~\cite{nguyen2022digital}}&-&TaS&IoT&Railway&ITS&Simulation%Intelligent Transport Systems 
            \\%\cmidrule(ll){2-4}
            &\textbf{~\cite{munoz2021using,munoz2022using,munoz2022measuring}}&MDE&DTM&CPS&Automotive&Cars&DT fidelity
            \\%\cmidrule(ll){2-4}
            &\textbf{~\cite{david2021inference}}&RL&Python-PDEVS&IoT&DEVS&Agent-based&Simulations inference
            \\%\cmidrule(ll){2-4}
            &\textbf{~\cite{lehner2021aml4dt}}&ML&AML4DT&CPS/IoT&Control&Air quality&Maintenance
            \\%\cmidrule(ll){2-4}
            &\textbf{~\cite{elayan2021digital}}&ML&Nonpublic&IoT&Healthcare&ECG&Predictions
            \\
		\cmidrule(ll){1-8}
            \multirow{11}{*}{\textbf{\rotatebox[origin=c]{90}{\makecell{Testing IoT Systems}}}}
            &\textbf{~\cite{li2022domain}}&DSL&IoTECS&IoT&Edge&-&Simulation %testing
            \\%\cmidrule(ll){2-4}
            &\textbf{~\cite{gupta2020towards}}&Architecture&HIPA&IoT&Security&-&Handle privacy violations
            \\%\cmidrule(ll){2-4}
            &\textbf{~\cite{gutierrez2019evolutionary}}&Search-based&IoT-TEG&IoT&Cross-domain&-&Mutation testing
            \\%\cmidrule(ll){2-4}
            &\textbf{~\cite{kirchhof2022model}}&MDE&Monti-Things&IoT&Devices&-&Cost-effective testing
            \\%\cmidrule(ll){2-4}
            &\textbf{~\cite{gupta2021hierarchical}}&ML&Nonpublic&IoT&Healthcare&PatientsDT&Anomaly detection
            \\%\cmidrule(ll){2-4}
            &\textbf{~\cite{de2023digital}}&Architecture&Public&CPS/IoT&Railway&General&Anomaly detection
            \\%\cmidrule(ll){2-4}
            &\textbf{~\cite{xu2021digital,xu2023digital}}&ML&N/A&CPS&Cross-domain&General&Anomaly detection
            \\
		\cmidrule(ll){1-8}
            \multirow{4}{*}{\textbf{\rotatebox[origin=c]{90}{\makecell{EM}}}}
            &\textbf{~\cite{moawad2015beyond}}&MDE&KMF&IoT&Smart systems&-&Handle volatile data
            % \\%\cmidrule(ll){2-4}
            \\%\cmidrule(ll){2-4}
            &\textbf{~\cite{leduc2017revisiting}}&DSL&ALE&General&General&-&Executable models
            \\%\cmidrule(ll){2-4}
            &\textbf{~\cite{riccobene2020model,bonfanti2022compositional}}&MDE&ASMETA&General&General&-&ASM Simulation
            % \\%\cmidrule(ll){2-4}
            \\
        \cmidrule(ll){1-8}
            \multirow{3}{*}{\textbf{\rotatebox[origin=c]{90}{\makecell{Ind.}}}}
            &\textbf{~\cite{iothub,azuredt}}&-&Azure&CPS/IoT&General&General&General
            % \\%\cmidrule(ll){2-4}
            \\%\cmidrule(ll){2-4}
            &\textbf{~\cite{iottwinmaker}}&-&AWS&IoT&General&General&General
            \\%\cmidrule(ll){2-4}
            &\textbf{~\cite{vorto,hono,ditto}}&-&Eclipse&CPS/IoT&General&General&General
            % \\%\cmidrule(ll){2-4}
            \\
        \cmidrule(ll){1-8}
            \multirow{1}{*}{\textbf{\rotatebox[origin=c]{0}{\makecell{Our Work}}}}
            &\textbf{}&MDE&APD-DT~\cite{repo}&IoT&Healthcare&Medicine Dispenser&Automated Testing
            \\
		\bottomrule
    \multicolumn{8}{l}{\fontsize{7}{8}\selectfont MDE: Model-Driven Engineering, DEVS: Discrete Event System Speciﬁcation, Ind.: Industrial, EM: Executable Models,}\\
    \multicolumn{8}{l}{\fontsize{7}{8}\selectfont DSL: Domain-Specific Language, RL: Reinforcement Learning, SW Dev.: Software Development, ECG: Electrocardiogram}\\
    \multicolumn{8}{l}{\fontsize{7}{8}\selectfont GUI: Graphical User Interface, ASM: Abstract State Machine, ITS: Intelligent Transportation System.}
	\end{tabular}
	\label{tab:rwcomp}
\end{table*}

\section{Related Work}\label{rws}
We relate our work with existing literature considering DT generation and operation, testing IoT systems, runtime/executable models, and industrial DTs. 
We consider these aspects relevant because our work targets DT generation and operation using executable models to support automated testing of industrial healthcare IoT applications. 
Table~\ref{tab:rwcomp} compares our work with each type of related work based on different functionalities, approach type, tool name/availability, target system and domain, DT category, and scope/purpose. Based on the overall comparison, our work contributes to the model-based development of DTs of medicine dispensers used with healthcare IoT applications. Following, we further analyze literature works and relate them with our work. 

\textbf{DT Generation and Operation.}
Several approaches utilize domain-specific modeling languages~\cite{dalibor2022generating,michael2021towards}, metamodels~\cite{pfeiffer2022modeling, corradini2023dtmn, yue2021understanding}, and model-driven techniques~\cite{bordeleau2020towards, govindasamy2021air} for creating DTs in various domains such as DTs for managing graphical user interactions~\cite{bano2022process, michael2022generating, dalibor2020towards}, satellite systems~\cite{christofi2022novel}, and web applications~\cite{bonney2021digital}. 
Some approaches are dedicated to creating DTs for cyber-physical systems (CPSs), such as~\cite{barat2022digital, stary2022behavior, kirchhof2020model}.
Similar to these works, our approach uses MDE to generate and operate DTs. However, our approach targets DTs for medicine dispensers to enable automated system-level testing, which is different from existing works.

Some techniques use machine learning (ML) for creating DTs with different purposes, such as for learning simulations of a physical entity~\cite{david2021inference} in IoT system, for creating and maintaining generic CPSs/IoTs~\cite{lehner2021aml4dt}, and for predicting heart conditions~\cite{elayan2021digital}.  
Our approach does not require patients' medical data, which is a primary requirement for creating ML-based DTs. 
A few works, such as ~\cite{kirchhof2021understanding, nguyen2022digital}, use DTs for simulating IoT clouds. 
Compared to these works, our approach targets a different purpose: generating and operating DTs for replacing concurrently operating medicine dispensers without requiring patient data for training ML models.

Some works focus on validating DTs~\cite{munoz2021using} and assessing the fidelity of DTs~\cite{munoz2022using, munoz2022measuring}. 
Inspired by these works, we also evaluated the fidelity of DTs generated with our approach. 
A few studies were conducted to analyze techniques for constructing DTs ~\cite{segovia2022design}, the industrial lifecycle of DTs~\cite{kaur2023standards}, and virtual reality and gamification using DTs~\cite{bucchiarone2022gamification}. 
These works mainly focus on devising an approach to analyze DT, whereas our work mainly targets creating and operating DTs.  
Similar to these works, we assess DTs' fidelity based on functional similarity using the method presented in~\cite{munoz2021using}. However, the difference is that the DT category, in our case, is medicine dispensers.

\textbf{Testing IoT Systems.} 
There are approaches for testing IoT applications~\cite{li2022domain}, handle privacy violations in IoT clouds~\cite{gupta2020towards}, mutation testing of IoT applications~\cite{gutierrez2019evolutionary}, and 
cost-effective deployment and testing of IoT clouds ~\cite{kirchhof2022model}.
Some approaches utilize DTs for anomaly detection in healthcare IoT systems~\cite{gupta2021hierarchical}, industrial IoT systems~\cite{de2023digital}, and CPSs~\cite{xu2021digital,xu2023digital}. 
Compared to these works, our approach aims to facilitate automated testing of healthcare IoT applications using DTs in place of medical dispensers. The key difference is that our work does not focus on devising a testing approach for IoT applications. Thus, some existing testing techniques could be applied in our case. However, which testing technique will work the best in our real-world context requires an investigation of its own.

\textbf{Executable Models.}
Approaches are available for creating executable models~\cite{bencomo2019models}, using such as 
models@run.time~\cite{moawad2015beyond}, domain-specific modeling languages~\cite{leduc2017revisiting}, and abstract state machines at run-time~\cite{riccobene2020model, bonfanti2022compositional}. 
These works provide generic/domain-specific approaches for runtime/executable models. Our approach utilizes executable models to construct DT behavior, which is a part of the overall process for creating and operating DTs of medicine dispensers. The focus of work is a main differentiating point.

\textbf{Industrial DTs.} Many industrial solutions are available for developing DTs, such as Azure IoT Hub~\cite{iothub} with Azure Digital Twins~\cite{azuredt} service, AWS IoT TwinMaker~\cite{iottwinmaker}, Vorto~\cite{vorto}, Hono~\cite{hono} and Ditto~\cite{ditto} from Eclipse. 
One of their key limitations is restricting the number of requests, which is unsuitable for conducting rigorous testing. 
Also, the primary limitation of open-source tools is public data storage~\cite{eclipseterms}, which causes privacy issues for healthcare authorities. 
Further, they are highly generic~\cite{dalibor2022cross}, and application-specific functionalities need to be added manually~\cite{lehner2022digital,pfeiffer2022modeling}.
Moreover, some commercial tools (e.g., AWS IoT TwinMaker) restrict their use in the healthcare domain due to their safety-critical nature. 
In comparison, our work provides a domain-specific DT generation approach to facilitate rigorous and automated testing of healthcare IoT applications. Also, our work does not involve data privacy concerns, which will be the focus of our future work. 

\section{Conclusion}\label{conclusion}
Automated testing of IoT-based healthcare applications is essential to ensure their intended functionalities are dependable. However, in a real-world setting, such applications include many medical devices from various vendors, varied cloud infrastructures, and software versions. Thus, performing testing at a large scale with many physical devices is practically impossible.  
To this end, we proposed a model-based approach to create and operate Digital Twins (DTs) of physical devices. We focused on the real context of Oslo City and its industrial partner delivering industrial medicine dispensers (Karie). 
We analyzed the fidelity of the DT developed with our approach by comparing functional similarity considering request response times and status codes with Karie's request responses. Results showed that the DT behaved more than 92\%, similar to Karie. We also successfully demonstrated that it is scalable to integrate 100 DTs (simulating 100 physical devices operating concurrently) into the test infrastructure.
In the future, we plan to assess the DT's fidelity by integrating diverse types of devices, performing a large-scale evaluation with DTs representing thousands of physical devices, and enriching the device domain model with concepts to encompass a variety of medical devices, such as health monitoring devices.

\section*{Acknowledgments}
This research work is a part of the WTT4Oslo project (No. 309175) funded by the Research Council of Norway. All the experiments reported in this paper are conducted in a laboratory setting of Simula Research Laboratory; therefore, they do not by any means reflect the quality of services Oslo City provides to its citizens or the quality of Oslo City's healthcare products.

\bibliography{refs}

\end{document}